\documentclass[preprint,amsmath,amssymb,aps]{revtex4-2}
\usepackage{graphicx,siunitx}       % figure files
\usepackage{dcolumn, booktabs}      % Align table columns on decimal point
\usepackage{bm,lipsum,tabularx}    % bold math

\usepackage{xcolor,ifthen}
\newboolean{showred}
\newboolean{showblue}
\setboolean{showred}{false}   % turn red ON or OFF
\setboolean{showblue}{false} % turn blue ON or OFF
\newcommand{\highlightred}[1]{%
  \ifthenelse{\boolean{showred}}{\textcolor{red}{#1}}{#1}}
\newcommand{\highlightblue}[1]{%
  \ifthenelse{\boolean{showblue}}{\textcolor{blue}{#1}}{#1}}

\usepackage[pdfpagelabels, pdfencoding=auto, psdextra]{hyperref}
\hypersetup{%
 pdfsubject=Paper,
 pdfkeywords={nuclear physics} 
 unicode = true,
 breaklinks = true,
 colorlinks = true,
 linkcolor = blue,
 menucolor = blue,
 citecolor = blue,
 urlcolor = blue
}

\begin{document}
\title{Photolithography-Compatible Three-Terminal Superconducting Switch for Driving CMOS Loads}

\author{Dip Joti Paul}
\email{djpaul@mit.edu}

\author{Tony X. Zhou}
\altaffiliation{Present address: Northrop Grumman Mission Systems, Linthicum, Maryland 21090, USA} 

\author{Karl K. Berggren} 
\email{berggren@mit.edu}
\affiliation{Research Laboratory of Electronics, Massachusetts Institute of Technology, Cambridge, MA 02139, USA}
% \date{\today}

\begin{abstract}
Superconducting devices have enabled breakthrough performance in quantum sensing and ultra-low-power computing. Nevertheless, the need for a cryo-electronics platform that can interface superconductor electronics with Complementary Metal-Oxide-Semiconductor (CMOS) devices has become increasingly evident in many cutting-edge applications. In this work, we present a three-terminal micrometer-wide superconducting wire-based cryotron switch (wTron), fabricated using photolithography, that can directly interface with CMOS electronics. The wTron features an output impedance exceeding 1 k$\Omega$ and exhibits reduced sensitivity to ambient magnetic noise, similar to its nanoscale predecessor, the nanocryotron. In addition, its micrometer-wide wires support switching currents in the mA range, making wTrons well-suited for driving current-hungry resistive loads and highly capacitive CMOS loads. We demonstrate this capability by using the wTron to drive room-temperature CMOS electronics, including an LED and a MOSFET with a gate capacitance of 500 pF. We then examine the optimal design parameters of wTrons to drive CMOS loads, such as MOSFETs, HEMTs, and electro-optic modulators. \highlightblue{Furthermore, to demonstrate the foundry readiness of the wTron, we fabricated wTrons using MIT Lincoln Laboratory’s SFQ5ee superconducting process and characterized their switching behavior.} Our work shows that wTron will facilitate the interface between superconductor electronics and CMOS, thereby paving the way for the development of foundry-compatible cryo-electronic ecosystems to advance next-generation computing and quantum applications.
\end{abstract}

\maketitle

\vspace{-6mm}
\section{Introduction}
\vspace{-3mm}
Superconductor electronics (SCE) is a promising and key area of interest for beyond-CMOS energy-efficient computing, cryogenic sensor readout, and quantum computing platforms \cite{R1,R2}. Research over the past several decades has led to the development of a wide array of superconducting switching devices based on the Josephson effect \cite{R3}. However, a critical yet often overlooked challenge remains persistent in the SCE platform, in particular the demonstration of a robust on-chip interface between SCE and CMOS electronics. For SCE to thrive in more practical applications, addressing this CMOS compatibility challenge is essential. While capable of operating at GHz clock frequencies with sub-aJ/bit energy dissipation, Josephson junction (JJ)-based switches are less ideal for applications that require mA-range currents to drive large resistive or capacitive impedances. As a result, the integration of JJ-based electronics into the CMOS platform remains elusive to this date.

Reviving a 1960s concept called cryotron \cite{R4}, a three-terminal superconducting nanowire-based switch was fabricated on-chip in 2014 \cite{R9} and has shown promising potential to bridge the gap between the SCE and CMOS platforms \cite{R13}. This family of superconducting nanowire-based electronics, including the nanocryotron (nTron) and heater nanocryotron (hTron) \cite{R9,R10}, builds on the nanofabrication techniques used for superconducting nanowire single-photon detectors (SNSPDs). Two decades after the discovery of the nanowire-based single-photon detector, research breakthroughs have enabled micrometer-wide wires sensitive to single photons \cite{R18}. However, micrometer-wide cryotrons have yet to receive attention, as cryotrons are still fabricated at sub-100 nm dimensions using electron-beam lithography for monolithic integration with SNSPDs. 

\highlightblue{Although there is no fundamental physical limitation that would inhibit superconducting wire-based cryotrons from operating at micrometer widths, their development has primarily been limited by a lack of application-driven motivation. The early development of nanowire-based cryotrons, such as the nTron and hTron, was closely tied to SNSPD readout applications \cite{R9,R10}. Consequently, cryotron designs remained at nanometer-scale dimensions to maintain compatibility with SNSPD fabrication processes and to support low-energy, high-speed digital logic circuits rather than high-current drive capability. In addition, increasing the wire width usually leads to higher switching energy and potentially slower operating speed, which has been viewed as a disadvantage in applications where speed and energy efficiency are prioritized over delivering large output currents to the load. Thus, the performance trade-offs, coupled with the lack of demand for high-current-driving superconducting switches at the time, have limited the exploration of micrometer-wide cryotrons (wTrons). In this work, we demonstrate that wTrons are not only feasible to fabricate using photolithography, but also advantageous for driving larger capacitive or resistive loads, making them well-suited for CMOS interfacing applications.}

The micrometer-wide cryotron (wTron) offers several advantages over nTron while retaining the benefits of nTron compared to JJ-based circuits, such as reduced sensitivity to stray magnetic fields and k$\Omega$-range output impedance \cite{R9}. Since the switching current of wTrons is in the mA range and their output impedance is in the k$\Omega$ range, they can drive high currents to load impedances. In addition, while the fabrication of nTrons requires an electron-beam lithography system, wTrons can be conveniently fabricated using photolithography, making foundry process integration more feasible. The micrometer-wide features of wTrons, compared to the sub-100 nm dimensions of nTrons, exhibit higher yield and better uniformity in widths since the concern of non-uniformities or constrictions along the wires is less pronounced for wider wires. This leads to fewer variations in switching currents across devices with similar geometry, demonstrating enhanced operating margins in wTron-based devices. Moreover, this photolithography-compatible cryotron can be readily fabricated using the commercial superconducting foundry process \cite{mitll_superconducting, tolpygo2016advanced}, facilitating on-chip integration of cryotron devices with JJ-based electronics and other superconducting devices \cite{R30}. This on-chip integration has the potential to enable complex functionalities by leveraging the strengths of both JJ and wTron devices while interfacing with CMOS, which could be useful for a broader range of practical applications. 

% While wTrons offer several advantages, they may present some trade-offs compared to nTrons, including higher switching energy, slower reset times, and lower current gain. More concrete comparisons of these performance metrics, however, require further characterization. Nevertheless, wTrons may not be ideal for applications where nTrons excel, such as those requiring high small-signal amplification gain or sensitivity to low-amplitude input signals from SNSPDs or SFQs. Hence, in certain applications, wTrons can serve as a complementary alternative to nTrons, and their combined use may offer a more versatile solution by leveraging the strengths of both devices across a broader range of applications.

In this work, we investigated the characteristics of wTrons by fabricating several devices with varying choke and channel widths using photolithography. We then performed DC I-V measurements to characterize their switching behavior at varying gate and channel bias currents. Next, we demonstrated the ability of wTrons to drive semiconductor devices by dynamically controlling the on/off state of an LED and a MOSFET. In addition, we developed a circuit model of the wTron to optimize its capacitive load-driving performance at MHz clock frequencies. This modeling framework will be useful for circuit design and performance optimization when integrating wTrons with functional CMOS and JJ-based circuits. \highlightblue{Finally, we demonstrated the foundry compatibility of the wTrons by fabricating several devices using MIT Lincoln Laboratory’s SFQ5ee process and performing electrical characterization, thus showing the feasibility of their monolithic integration with JJ-based electronics.}

\vspace{-6mm}
\section{Methods}
\vspace{-3mm}
\subsection{Device fabrication}
\vspace{-3mm}
The devices characterized in this work were fabricated using photolithography. The fabrication flow is as follows: First, we deposited 10-nm-thick NbN film on 300-nm-thick oxidized silicon substrate at room temperature using an AJA Orion magnetron sputtering system in a cryopumped chamber at base pressure of 5$\times 10^{-9}$ Torr \cite{R26}. The sheet resistance of the deposited film was measured to be 251.65 $\Omega/\square$ at room temperature, with a critical temperature ($T_\mathit{c}$) of 8.9 K. Next, the NbN sample was spin-coated with a positive-tone resist (AZ 3312) at 3000 rpm for 60 s and then \highlightblue{baked on a hotplate at 110$^{\circ}$ C for 90 s.} The sample was then exposed to 405 nm wavelength light using a Heidelberg MLA 150 maskless aligner, followed by development in AZ 726 at room temperature for 60 s, and rinsed in DI water for 10 s. Following this, reactive ion etching was performed on the sample with \si{CF_4} plasma at 30 W. The etching process occurred in two steps: 2 minutes of etching in each step with a 1-minute interval between the two steps. This prevented the resist layer from overheating and hardening during etching. Finally, the resist lift-off was performed by immersing the sample in Microposit Remover (N-Methyl-2-pyrrolidone) solvent heated at 60$^{\circ}$ C for 1 hour. The smallest achievable feature width using this photolithography process was 1 $\mu$m.

A scanning electron micrograph (SEM) of a wTron is shown in Figure \ref{fig:Fig1}(a). The choke and channel are gradually tapered, and the narrowest widths of the choke and channel define their switching currents. When the choke switches, the resulting Joule heating from its tapered geometry locally increases the temperature in the channel wire, thereby suppressing its critical current. In addition, Figure \ref{fig:Fig1}(a) shows that the source and drain terminals in the wTron can be interchanged due to their geometric symmetry.

\vspace{-6mm}
\section{Results}
\vspace{-3mm}
\subsection{DC I-V measurements of \lowercase{w}T\lowercase{rons}} 
\vspace{-3mm}
The wTron switches from the superconducting to the resistive state when the current through the channel wire exceeds its switching current. However, due to the spatial proximity of the choke and channel regions, a resistive hotspot formed in the choke suppresses the switching current of the channel wire. The extent of this suppression varies depending on factors such as the thermal conductivity of the superconducting material and substrate, as well as the widths of the choke and channel wires. This modulation of the channel switching current by applying a gate current is also observed in nTron devices \cite{R9, R12}.

\begin{figure}[!t]
\centering
\vspace{-3mm}
\includegraphics[width=0.83\linewidth]{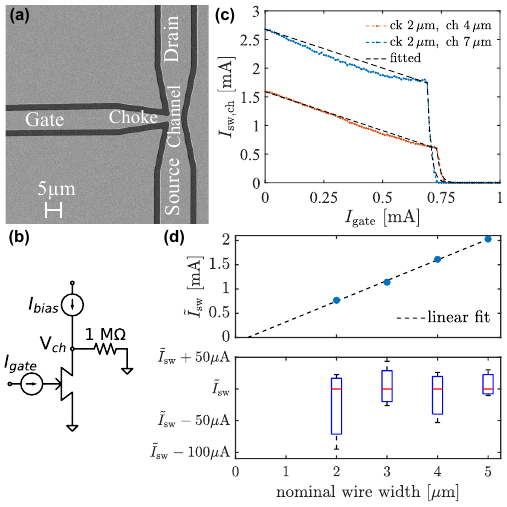}
\vspace{-5mm}
\caption{\footnotesize (a) A scanning electron micrograph (SEM) of a fabricated wTron is shown, with a 2 $\mu$m-wide choke (ck) and 4 $\mu$m-wide channel (ch). (b) A schematic of the circuit used for the DC I–V measurement of the wTron is shown. Yokogawa GS200 was used as the current source, and the voltage across the channel ($V_{ch}$) was measured to identify the switching state of the wTron. Here, 1 M$\Omega$ denotes the internal resistance of the voltmeter. (c) Channel switching current ($I_{sw,ch}$) is plotted as a function of gate current for two different choke and channel widths of the wTron (blue/red: experimental data; black: fitted data). Measurements were performed at 3.8 K. (d) The upper panel shows the median switching currents ($\tilde{I}_{sw}$) of several photolitho-fabricated micrometer-wide wires plotted as a function of wire width, with five wires measured for each nominal width. The box plot in the lower panel shows the corresponding switching current distributions. The top and bottom edges of each box represent the upper and lower quartiles, respectively, and the red line indicates the median of the distribution. \vspace{-3mm}}
\label{fig:Fig1}
\end{figure} 

To characterize this behavior in micrometer-wide cryotrons, we performed DC I–V measurements on two sets of wTrons with different choke and channel widths. \highlightblue{A circuit schematic for the DC I–V measurement of the wTron is shown in Figure \ref{fig:Fig1}(b). The channel bias current ($I_{bias}$) was gradually increased at a fixed gate current ($I_{gate}$), while the channel voltage ($V_{ch}$) was monitored to identify the onset of resistance in the channel, which allowed us to determine the channel switching current ($I_{sw,ch}$) at that gate current. This process was repeated for gate current ranging from zero to above the choke switching current, with fine increments and sufficient delay intervals to ensure the device returned to the superconducting state after each switching event.} 

Figure \ref{fig:Fig1}(c) shows the channel switching current ($I_{sw,ch}$) as a function of the gate current. The choke width in both wTrons was 2 $\mu$m, and the choke switching current ($I^{0}_{sw,ck}$) was around 0.7 mA. In both wTrons, due to the summation of $I_{gate}$ and $I_{bias}$ at the choke-channel junction, $I_{sw,ch}$ decreased almost linearly with $I_{gate}$ until it reached the choke switching current. Then, $I_{sw,ch}$ sharply decreased as $I_{gate}$ increased further, due to the resistive hotspot in the choke suppressing the superconductivity of the channel. This suppression of channel switching current resulting from $I_{gate}$ was empirically fitted to the following equation:
\begin{equation}
I_{sw,ch} = \begin{cases} 
I^{0}_{sw,ch} - \alpha I_{gate}   &\mbox{if } I_{gate} < I^{0}_{sw,ck} \\
I^{0} \times \exp\left[-\beta \left(I_{gate} - I^{0}_{sw,ck}\right)\right] &\mbox{if } I_{gate} \geq I^{0}_{sw,ck} 
\end{cases} 
\label{eq:eq1}
\end{equation}
where $I^{0} = I^{0}_{sw,ch} - \alpha I^{0}_{sw,ck}$. \highlightblue{The notations $I^{0}_{sw,ch}$ and $I^{0}_{sw,ck}$ refer to the switching currents of the channel and choke wires, respectively, measured under isolated conditions, i.e., when current is applied to only one wire (either the channel or the choke), with no current applied to the other. Thus, these values represent the intrinsic switching current of the channel (or choke) wire in the absence of any influence from current applied to the other wire. The switching current of a micrometer-wide wire can be determined from its DC I-V measurements, as shown in Figure \ref{fig:FigA1} in Appendix A.}

The empirically determined values of the fitting parameters $\alpha$ and $\beta$ are 1.36 and $6.5 \times 10^{4}$, respectively. The same values of $\alpha$ and $\beta$ were used for both dimensions of the wTron in Figure \ref{fig:Fig1}(c), and both of the fitted lines are in good agreement with the experimental data. Although a linear fitting function could be used in the regime where $I_{gate} \geq I^{0}_{sw,ck}$, an exponential fitting function was chosen to represent the complex interplay of Joule heating-induced suppression of the channel switching current. 

We hypothesize that the value of $\alpha$ greater than 1 arises primarily from current crowding at the sharp edges of the choke-channel interface. Therefore, we define $\alpha$ as a measure of the current crowding factor in wTrons. As seen in Figure \ref{fig:Fig1}(c), discrepancies between the fitted curve and the experimental data at some $I_{gate}$ values in the linear region suggest that $\alpha$ may vary depending on the bias current, geometry, and edge roughness of the choke-channel interface. In addition, the parameter $\beta$ represents hotspot-induced local suppression of superconductivity and is expected to depend on factors such as the thermal conductivity of the superconducting material and substrate, as well as the dimensions of the choke and channel widths. The value of $\beta$ was determined by adjusting it during the curve-fitting process until a satisfactory match with the experimental data was observed.

\highlightblue{We acknowledge that quantum and thermal fluctuations, such as vortex edge entry or vortex–antivortex pair unbinding, may also contribute to the suppression of the channel switching current. However, incorporating these mechanisms into the device model to better fit the experimental data would require a more detailed thermal and quantum modeling framework, which is beyond the scope of the present work. Instead, the empirical model presented here was intended to capture the macroscopic switching behavior of wTrons and to support the development of an electrical circuit model of the wTron for use in superconductor electronics \cite{R25}, such as capacitive load-driving circuits. In this work, we implemented the wTron circuit model in LTspice, a SPICE-based electronic circuit simulation tool \cite{R34}, and compared the simulation results from the developed SPICE model of the wTron with experimental data, as detailed in Appendix A.}

% \highlightblue{Finally, building on this developed empirical Equation \ref{eq:eq1}, which models the channel switching current of a wTron as a function of its gate current, an electrical circuit model of the wTron can be developed. In this work, we developed the electrical circuit model of the wTron using the SPICE-based electronic circuit simulation software LTspice \cite{R34}, as discussed in detail in Appendix A.}

% We note that the value of $\alpha$ is greater than 1, indicating an additional factor contributing to current suppression beyond the summation of currents at the choke-channel junction. We hypothesize that this arises from current crowding at the sharp edges of the choke-channel interface. Therefore, we define $\alpha$ as a measure of the current crowding factor in wTrons. As seen in Figure \ref{fig:Fig1}(c), there are discrepancies between the fitted curve and the actual data at some $I_{gate}$ values in the linear section, suggesting a higher value of $\alpha$ than 1.36 at those biases. We speculate that the value of $\alpha$ may vary depending on the shape and edge roughness of the choke-channel interface. 

Next, we performed DC I-V measurements of several photolitho-fabricated micrometer-wide wires to quantify the spread of their switching currents. The wire widths ranged from 2 $\mu$m to 5 $\mu$m, with each wire's length being 50 times its width. We measured five wires for each width at 3.8 K, and their switching current distribution is shown in Figure \ref{fig:Fig1}(d). As observed, the median value of the $I_{sw}$ distribution increased linearly with wire width, and the switching current density was calculated to be approximately 40 \si{GA/m^2}. Additionally, the switching current exhibited a linear dependence on wire width, given that the wire width was significantly smaller than the Pearl length in our device. The Pearl length was calculated as $\Lambda_P = 2\lambda^2/d = 42.3$ $\mu$m, where $d$ is the film thickness and the magnetic penetration depth ($\lambda$) of 10 nm thick NbN film is 460 nm \cite{R22}. The standard deviation of the switching current was approximately 100 $\mu$A or below, as shown in Figure \ref{fig:Fig1}(d). While stochastic and thermal fluctuations contribute to the uncertainty in the switching current of the superconducting wire \cite{R21}, we anticipate that lithography-induced line-edge roughness was likely the primary contributor to this observed $I_{sw}$ variation. Although the smallest feature fabricable with our photo-litho process was 1 $\mu$m, the process was not optimized enough to yield smooth line edge roughness at that minimum feature size. Further improvements in our photolithography process are possible, which may decrease the spread of switching current in the micrometer-wide wires and, thus, reduce the switching uncertainty in wTron devices.

\highlightblue{The wTron, similar to the nTron, primarily functions as a digital comparator and does not actively amplify the input signal. However, by forming a resistive region in the channel with a small gate current $I_{gate}$, a fraction of the bias current ($I_{bias}$) can be directed to the load impedance, a phenomenon referred to as current gain in superconducting wire-based cryotron devices \cite{R9}. The upper bound of the current gain, for a given load impedance, can be estimated from the ratio of the output load current swing to the gray zone width of the choke switching current (i.e., the change in gate current required to trigger switching in the choke) when the gate is biased close to the choke switching current \cite{R31}. The gain of a cryotron device can vary depending on operating conditions such as load impedance, sheet resistance, critical current density of the superconducting film, and operating temperature. In nTron devices with sub-100 nm choke widths, the gain typically ranges from 20 to 250 \cite{R9, R31}. To quantify the gain of the wTron at a given load impedance, we need to measure the switching current distribution of the micrometer-wide choke and the change in output current delivered to the load due to the switching of the channel. Although we measured the switching current distribution of several photolitho-fabricated wires to be around 100 $\mu$A, the observed distribution was primarily influenced by lithography-induced linewidth variations across different wires and therefore did not provide the gray zone width of the switching current for a micrometer-wide choke. While a quantitative gain estimation of micrometer-wide cryotron devices was not performed in this work, it will be important to investigate in future work to optimize the wTron device parameters for enhanced load-driving capability.}

% However, we estimate that the gain of the wTron will fall within the typical range of the nTron, as the stochastic fluctuation-induced gray zone width of the micrometer-scale choke is expected to be comparable to the nanometer-scale choke in the nTron \cite{R31}.

% \subsection{Driving room-temperature LED and MOSFET with \lowercase{w}T\lowercase{ron}}
\vspace{-6mm}
\subsection{Demonstration of driving CMOS loads with \lowercase{w}T\lowercase{ron}}
\begin{figure}[!t]
\centering
\vspace{-3mm}
\includegraphics[width=0.8\linewidth]{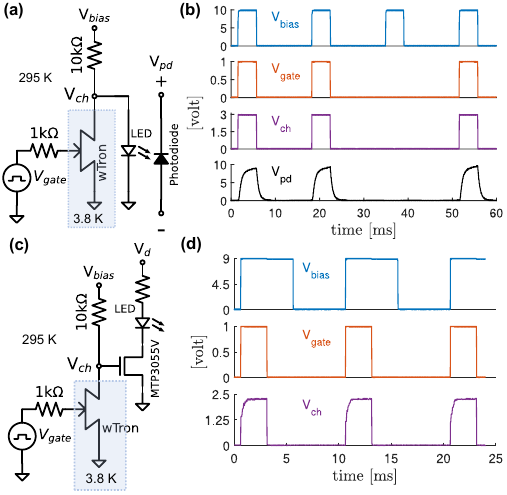}
\vspace{-5mm}
\caption{\footnotesize (a) A schematic of driving an LED with a wTron is shown. The LED was at room temperature and connected to the wTron channel ($V_{ch}$) using coaxial cables (not shown in the schematic). The wTron gate and channel were connected to a dual-channel arbitrary waveform generator in series with bias resistors. A low-noise DC voltage source was used to bias the photodiode, and an oscilloscope captured the output voltage traces. (b) When the wTron was switched, the channel output voltage ($V_{ch}$) exceeded the forward voltage of the LED. This turned on the LED, as indicated by the voltage across the photodiode ($V_{pd}$). (c) A schematic of driving a MOSFET using a wTron is shown. The MOSFET was at room temperature and connected to the wTron channel ($V_{ch}$) with coaxial cables (not shown in the schematic). An LED was connected in series with the drain of the MOSFET to detect its on/off state. (d) When the wTron was switched, the gate voltage ($V_{ch}$) of the MOSFET increased. The MOSFET turned on once $V_{ch}$ exceeded its gate threshold voltage of 2.1 V. When $V_{gate}$ returned to zero, the wTron reverted to its superconducting state, with turning off the MOSFET. \vspace{-3mm}}
\label{fig:Fig2}
\end{figure}

\vspace{-3mm}
In this section, we demonstrate the use of the wTron to drive CMOS electronics, such as an LED and a MOSFET. Interfacing with cryogenic low-power LEDs has previously been demonstrated using nTrons \cite{R19}. In contrast, in this work, we used a relatively high-power blue LED with a forward voltage of 2.7 V and demonstrated gate-controlled switching of the LED using a wTron, as shown in Figure \ref{fig:Fig2}(a). A photodiode (Thorlabs DET100A) was placed close to the LED using an optical mount to monitor the on-state of the LED electrically. The voltage across the LED needed to be at least 2.9 V to produce a detectable light intensity for the photodiode used in this experiment. This required delivering 0.3 mA of current to the LED, resulting in 0.9 mW power dissipation in the LED. In this experiment, the choke and channel widths of the wTron were 2 $\mu$m and 4 $\mu$m, respectively. As shown in Figures \ref{fig:Fig2}(a) and \ref{fig:Fig2}(b), the wTron channel was biased with 10 V voltage source in series with a 10 k$\Omega$ resistor, resulting in 1 mA current through the channel wire. However, since the switching current of the 4 $\mu$m wide channel was 1.6 mA, no voltage drop across the channel ($V_{ch}$) was observed when the gate input ($V_{gate}$) was zero. But when a gate voltage of 1 V was applied in series with a 1 k$\Omega$ resistance, the 2 $\mu$m choke was switched, as its switching current was 0.75 mA. The switching of the choke suppressed the channel critical current, causing the wTron channel to switch at the given channel bias current. This, in turn, led to a high channel impedance, directing most of the channel bias current into the LED. The on/off state of the LED was identified from the output voltage ($V_{pd}$) of the photodiode. In addition, we observed that the luminous intensity of the LED increased with $V_{bias}$ as more power was delivered to it. 

\highlightblue{In this measurement, the wTron latched into a resistive state upon switching and maintained the LED’s on-state until a zero-bias voltage was applied to actively reset the wTron channel. The recovery of the superconducting state after switching to the resistive state in superconducting wire-based devices, such as SNSPDs and nTrons, depends on the trade-off between the time required for thermal cooling of the resistive region and the time needed to restore current from the load to the device \cite{R20}. The thermal relaxation time, influenced by the properties of the superconducting film and substrate, device linewidth, bias current, and operating temperature, is on the order of 100 ps for 10 nm NbN film on \si{SiO_2}/Si substrate \cite{R33}. The current dynamics in the wTron is determined by its inductive time constant, defined as $\tau = L_{ch}/R_{L}$, where $L_{ch}$ is the inductance between the channel and drain terminals, and $R_{L}$ is the load impedance \cite{R20}. The wTron used in this experiment had a channel-to-drain length of 50 squares, corresponding to channel inductance of 2 nH, assuming 40 pH/square kinetic inductance for the 10 nm-thick NbN film \cite{R26}. Since the load impedance of the LED in its on state was about 10 k$\Omega$, the inductive time constant ($\tau$) of the circuit was 0.2 ps, which is much smaller than the thermal relaxation time and thus insufficient to prevent the wTron from latching into a stable resistive state. The self-reset mode in the wTron can be enabled by adding additional inductance in series with the wTron channel, either by fabricating superconducting meanders or using external inductors.}

Next, we performed periodic switching of a MOSFET by applying the gate and channel currents to a wTron, as shown in Figure \ref{fig:Fig2}(c). In this experiment, we used an n-channel power MOSFET (MTP3055V), which had a gate threshold voltage of 2.1 V and an input capacitance of 500 pF. The choke and channel wire widths of the wTron were identical to the previous circuit in Figure \ref{fig:Fig2}(a). An LED was connected in series with the drain terminal of the MOSFET to identify its on/off state. As shown in Figure \ref{fig:Fig2}(d), the wTron gate and channel were biased with 1 mA and 0.9 mA amplitude pulses, respectively. This led to the switching of the wTron and the formation of several k$\Omega$ of resistance in its channel. Consequently, a fraction of the channel bias current was diverted to the gate of the MOSFET, which charged its gate-source and gate-drain capacitances. When the gate voltage of the MOSFET reached 2.1 V, the LED turned on, which indicated the on-state of the MOSFET. Figure \ref{fig:Fig2}(d) shows that the channel voltage ($V_{ch}$) initially ramped up exponentially, similar to the behavior observed during capacitor charging. The time taken to reach the gate threshold voltage of the MOSFET after applying the bias voltage is referred to as the turn-on delay, which depends on the channel hot-spot resistance of the wTron and the RC time constant of the circuit. However, since the MOSFET in this experiment was connected to the wTron via room-temperature coaxial cables, the load turn-on delay was also influenced by the distributed impedance of the cables. The load turn-on delay can be significantly reduced, and the output voltage across the capacitor can be increased by replacing the coaxial cables with low-impedance wire bonds. Nevertheless, this experiment demonstrates the viability of driving highly capacitive loads, such as cryo-CMOS and electro-optic modulators, with wTrons. For improved performance, however, the capacitive load should be interfaced on-chip with the wTron. In the following section, we discuss the design parameters of the wTron for driving such on-chip capacitive loads.

\vspace{-6mm}
\subsection{Design considerations for capacitive load driving with \lowercase{w}T\lowercase{ron}}
\vspace{-3mm}
Low-latency interfacing of cryogenic electronics with semiconducting and optoelectronic devices, such as MOSFETs, HEMTs, LEDs, and electro-optic modulators, holds the promise of revealing new possibilities for hybrid cryo-computing ecosystems. In contrast to JJ-based stacked amplifiers, cryotrons can provide a more convenient solution for integrating and interfacing with CMOS loads \cite{R13, R19}. In this section, we explore the design considerations to reduce the latency in driving CMOS loads with wTrons. In particular, we investigate the optimal dimensions of the choke and channel wire widths of wTrons for driving capacitive loads at MHz clock frequencies.

\begin{figure}[htbp!]
\centering
\vspace{-3mm}
\includegraphics[width=0.95\linewidth]{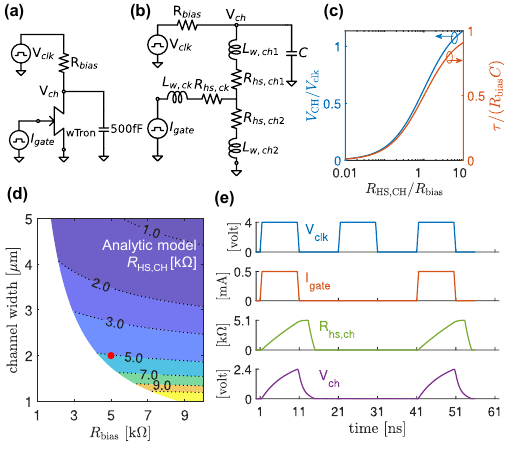}
\vspace{-5mm}
\caption{\footnotesize (a) Schematic of a wTron driving 500 fF capacitive load. (b) Equivalent circuit corresponding to (a), showing the instantaneous hot-spot resistance ($R_{hs}$) and series inductance ($L_w$) in the choke and channel regions of the wTron when it is in the resistive state. (c) The steady-state voltage across the capacitive load ($V_{CH}$) relative to the bias voltage amplitude ($V_{clk}$), and the normalized time constant of the circuit are plotted as functions of $R_{HS,CH}/R_{bias}$ when $I_{gate}$ = 0.5 mA. (d) The steady-state value of the channel hot-spot resistance ($R_{HS,CH}$) was calculated using an analytic model (see Appendix B, Equation \ref{eq:B7}) while sweeping both $R_{bias}$ and the channel wire widths. The calculation was performed for $V_{clk}$ frequency of 50 MHz with an amplitude of 4 V, and $I_{gate}$ amplitude of 0.5 mA. (e) LTspice simulation result of the circuit in (a), for $R_{bias}$ of 5 k$\Omega$, and wTron channel and choke width of 2 $\mu$m and 1 $\mu$m, respectively. The switching current of the choke is 0.46 mA; hence, $I_{gate}$-induced channel switching occurs in the wTron. Here, $R_{hs,ch}$ is the sum of $R_{hs,ch1}$ and $R_{hs,ch2}$, and the maximum steady-state value of $R_{hs,ch}$ matches the value computed using the analytic model (see Appendix B, Equation \ref{eq:B7}), as indicated by the red circle in (d). \vspace{-3mm}}
\label{fig:Fig3}
\end{figure} 

% The white region in the contour plots represents the pairs of $R_{bias}$ and channel width where the selected $R_{bias}$ results in a bias current ($V_{clk}/R_{bias}$) sufficient to switch the channel wire and, consequently, the wTron, without the need for $I_{gate}$. However, since $I_{gate}$ is intended to control the switching of the wTron and, consequently, the driving of the capacitor, the selection of these $R_{bias}$ and channel width pairs is not suitable for the intended operation.

In this study, we consider driving a 500 fF capacitor with a wTron, as shown in the schematic in Figure \ref{fig:Fig3}(a). This can be considered an equivalent circuit for driving a cryo-CMOS with wTron, where the gate capacitance is set to 500 fF. In the previous section, we demonstrated wTron driving a room-temperature n-channel power MOSFET having 500 pF gate capacitance. However, in this section, we consider driving a smaller capacitance as the typical capacitance of cryo-CMOS gates and electro-optic modulators falls within the range of 1 fF to 500 fF \cite{R37}. Although the gate threshold voltage of a cryo-CMOS is typically around 0.7 V, we aimed to design a wTron-based driving circuit capable of achieving 2 V across the 500 fF capacitive load. As the capacitor is connected in parallel with the wTron channel, the voltage across the capacitor appears when the wTron switches to the resistive state. The driving frequency of the capacitor is set by a clock bias ($V_{clk}$), which is connected in series with $R_{bias}$ to the wTron channel. The choke and channel widths of the wTron are chosen so that it switches to the resistive state only when a gate pulse ($I_{gate}$) is applied. Hence, $I_{gate}$ can be considered the control pulse for load driving. In addition, we considered a superconducting wire connecting the capacitor to the wTron; hence, we did not include interconnect impedance in the schematic of Figure \ref{fig:Fig3}(a).

To investigate how the output impedance of the wTron affects capacitive load driving, we developed an equivalent circuit shown in Figure \ref{fig:Fig3}(b), corresponding to the circuit in Figure \ref{fig:Fig3}(a) when the wTron is in the resistive state. In the resistive state, resistances form in the choke and channel regions of the wTron. These are often referred to as hot-spot resistances ($R_{hs}$), and their values vary over time depending on the current flowing through them \cite{R20, R25}. As shown in Figure \ref{fig:Fig1}(a), there is a geometric symmetry between the drain-channel and channel-source regions of the wTron, with the choke intersection point at the center. In Figure \ref{fig:Fig1}(b), we represent each of these three segments of the wTron with a resistance ($R_{hs}$) in series with an inductor ($L_w$). Here, the choke hot-spot resistance is denoted as $R_{hs,ck}$, while the hot-spot resistances of the upper and lower halves of the channel to the drain and source terminals are referred to as $R_{hs,ch1}$ and $R_{hs,ch2}$, respectively. In the resistive state of a superconducting wire, the series inductance ($L_w$) is primarily determined by the wire’s geometric dimensions, and this inductive impedance is generally smaller than the hot-spot resistance. However, in the superconducting state, the series inductance becomes the dominant impedance of the wire, while the hot-spot resistance becomes zero. In addition, we denote the load capacitor as $C$ and the bias resistance as $R_{bias}$. From the circuit in Figure \ref{fig:Fig1}(b), the maximum voltage across the capacitor ($V_{CH}$), i.e., the steady-state value after the capacitor is fully charged, can be calculated as follows:
\begin{equation}
V_{CH} = \frac{R_{HS,CH} + \eta R_{HS,CH2}}{R_{HS,CH}+R_{bias}} V_{clk}
\label{eq:eq2}
\end{equation}
Here, $R_{HS,CH}$ denotes the maximum (steady-state) value of the instantaneous hot-spot resistance of the channel ($R_{hs,ch}$), which is the sum of the hot-spot resistances in the upper ($R_{hs,ch1}$) and lower ($R_{hs,ch2}$) halves of the channel. Similarly, $R_{HS,CH2}$ represents the steady-state value of the instantaneous hot-spot resistance $R_{hs,ch2}$. Due to the symmetric drain-source geometry of the wTron, we assume $R_{HS,CH2}$ to be equal to $R_{HS,CH1}$, and thus each is half of $R_{HS,CH}$. The parameter $\eta$ denotes the ratio of $I_{gate}$ to $I_{bias}$, where $I_{bias}$ is the channel bias current defined as $V_{bias}/R_{bias}$. In Figure \ref{fig:Fig3}(c), we plot the value of $V_{CH}/V_{clk}$ as a function of $R_{HS,CH}/R_{bias}$ for $\eta = 0.5$, showing that a larger ratio of $R_{HS,CH}/R_{bias}$ results in a higher channel output voltage. This indicates that a high channel impedance ($R_{HS,CH}$) in the wTron is essential for effectively driving highly capacitive loads.

The RC time constant ($\tau$) of the circuit is given by $(R_{HS,CH} || R_{bias})C$, and $\tau$ increases with the increase of $R_{HS,CH}$ and $R_{bias}$ for a given load capacitance $C$, as shown in Figure \ref{fig:Fig3}(c). This indicates that when the control pulse ($I_{gate}$) is turned off, a larger value of $(R_{HS,CH} || R_{bias})$ results in a longer capacitor discharge time. This might cause the wTron to latch into a resistive state if the bias voltage ($V_{clk}$) is applied to the wTron channel before the capacitor fully discharges. Hence, the selection of $R_{bias}$ plays a role in setting the maximum driving frequency of the capacitive load. One can consider selecting a lower value of $R_{bias}$ to decrease the RC time constant of the circuit. However, a reduced $R_{bias}$ leads to a higher bias current in the wTron channel for a given $V_{clk}$ amplitude, which can cause the channel to switch even in the absence of the control pulse ($I_{gate}$). Since $I_{gate}$ is intended to control the switching of the wTron, and consequently, the charging of the capacitor, a wider channel is required to prevent unintended switching without $I_{gate}$. However, since the hot-spot growth rate of the channel is inversely proportional to its width, increasing the width will affect the steady-state value of the channel hot-spot resistance ($R_{HS,CH}$), thereby reducing the voltage across the capacitor. Therefore, optimizing the channel width and the bias resistance ($R_{bias}$) is required to achieve the target $V_{CH}$ while minimizing the RC time constant of the circuit.

Figure \ref{fig:Fig3}(d) shows the steady-state value of the channel hot-spot resistance as a function of channel width and bias resistance ($R_{bias}$). These values were calculated using an analytical expression (see Equation \ref{eq:B7} in Appendix B) derived from the electrothermal model of superconducting nanowires \cite{R20}. In Appendices A and B, we present the analytical model used in this work to estimate the hot-spot resistance and evaluate its validity for micrometer-wide wires. For the calculations, we assume the wTron is fabricated on a 10-nanometer-thick NbN film with a critical current density of 46 \si{GA/m^2} and a $T_{c}$ of 9 K. Square pulses of $V_{clk}$ and $I_{gate}$ with a frequency of 50 MHz and amplitudes of 4 V and 0.5 mA, respectively, are applied to the wTron channel and gate. As mentioned earlier, $I_{gate}$ is intended to control the on/off state of the wTron. Therefore, the channel width and $R_{bias}$ are chosen such that the bias current ($V_{clk}/R_{bias}$) switches the channel only when $I_{gate}$ is high. Figure \ref{fig:Fig3}(d) shows the valid ranges of $R_{bias}$ for different channel widths to ensure $I_{gate}$-controlled operation of the wTron, whereas values of $R_{bias}$ in the white region would cause the wTron channel to switch even without $I_{gate}$. As observed, the value of $R_{HS,CH}$ decreases for wider channel widths, and according to Equation \ref{eq:eq2}, this leads to a lower value of $V_{CH}$ for a given $V_{clk}$ amplitude. However, the RC time constant also decreases with the reduction in the wTron output impedance ($R_{HS,CH}$), as shown in Figure \ref{fig:Fig3}(c), enabling faster reset of the wTron and high-frequency operation of the capacitor.

Figure \ref{fig:Fig3}(e) shows the LTspice simulation results of the circuit, based on the SPICE electrothermal model of superconducting nanowires \cite{R25,R35}. The channel width and $R_{bias}$ were set to 2 $\mu$m and 5 k$\Omega$, respectively. As indicated by the red circle in Figure \ref{fig:Fig3}(d), the maximum steady-state value of the channel hot-spot resistance ($R_{HS,CH}$) predicted by our analytical model matches well with the LTspice result. Furthermore, we calculated the maximum value of $V_{CH}$ to be 2.64 V using Equation \ref{eq:eq2}, which also corresponds well with the LTspice simulation result in Figure \ref{fig:Fig3}(e). These results demonstrate the accuracy and usefulness of our analytical model in selecting the optimal channel width and $R_{bias}$ without requiring time-consuming iterative simulations or parameter sweeps in LTspice.

In summary, our analysis shows the feasibility of using the wTron for high-frequency driving of CMOS transistors and capacitive loads. The simulations show that a higher channel hot-spot resistance ($R_{HS,CH}$) results in a higher output voltage across the capacitor. Therefore, materials with higher sheet resistance are more suitable for fabricating cryotron-based circuits designed to drive capacitive loads. In addition, the RC time constant of the circuit should be shorter than the off-cycle duration of $V_{clk}$ to prevent the wTron from latching. Although a smaller value of $R_{bias}$ can reduce the RC time constant, it also leads to an increase in the bias current ($V_{clk}/R_{bias}$) to the channel. As a result, the channel width must be sufficiently large to ensure that its switching current exceeds the bias current, since we aim to operate the wTron in the regime where the channel switching is induced by $I_{gate}$. However, the hot-spot growth rate is inversely proportional to the wire width. Hence, the channel width affects the maximum value of the channel hot-spot resistance and, consequently, the voltage across the capacitor. Therefore, the value of $R_{bias}$ and the channel width need to be chosen optimally to achieve the desired voltage across the capacitor.

On the other hand, the requirement for selecting the choke width of the wTron is less stringent. Furthermore, as shown in Figure \ref{fig:Fig3}(b), when the wTron is in the superconducting state, the equivalent circuit becomes a parallel LC circuit, with $L_w$ representing the kinetic inductance of the wTron wires. These inductances result in a voltage ripple across the capacitor in the presence of the clock bias ($V_{clk}$), with the peak amplitude of the ripple voltage given by $V_{ripple} = (V_{clk}/R_{bias}) \sqrt{L/C}$. Improper choices of $R_{bias}$ and $L$ will result in a high $V_{ripple}$, thereby reducing the $V_{CH}/V_{ripple}$ ratio and potentially impacting the gate selectivity in the voltage control of the capacitor. This ripple amplitude can be kept below the design specification by appropriately choosing the total length of the channel wire, i.e., the channel kinetic inductance.

\vspace{-6mm}
\section{Foundry Compatibility of \lowercase{w}T\lowercase{ron} and I\lowercase{ts} Monolithic Integration with JJ-Based Electronics}
% \section{Foundry Compatibility and Monolithic Integration of \lowercase{w}T\lowercase{rons} with JJ-based Electronics}
\vspace{-3mm}
Superconducting nanowire-based cryotrons (nTrons and hTrons) have emerged as an alternative platform to JJ-based electronics due to their compatibility with monolithic fabrication processes, resilience to parasitic magnetic fields, high output impedance, high fan-out operation, and ability to drive high-impedance loads. To further advance their scalability and adoption in foundries and academic institutions with limited resources, we demonstrate the feasibility of fabricating superconducting wire-based cryotrons using standard photolithography. This compatibility will also enable on-chip integration with JJ-based electronics \cite{R3}, such as SFQ, RSFQ, and AQFP circuits, within superconducting foundry processes, thus opening up new possibilities for hybrid cryo-electronic systems.

\begin{figure}[!t]
\centering
\vspace{-3mm}
\includegraphics[width=0.95\linewidth]{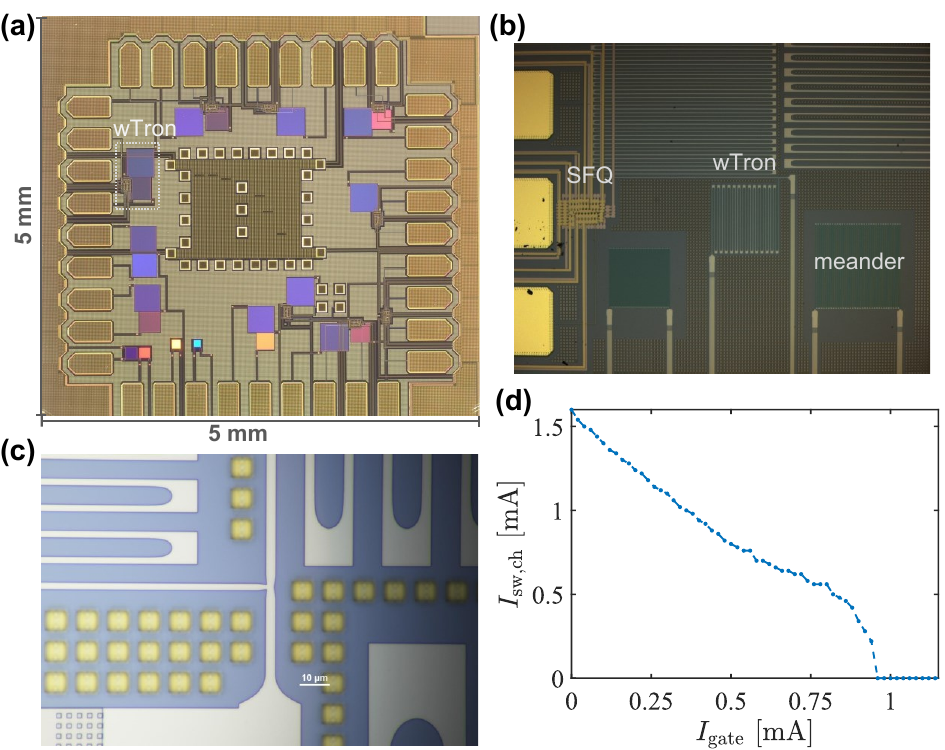}
\vspace{-5mm}
\caption{\footnotesize (a) An optical micrograph of a chip fabricated using MIT Lincoln Laboratory’s SFQ5ee foundry, consisting of multiple wTron devices, SFQ circuits, and superconducting wire meanders. (b) A segment of the chip showing a DC-to-SFQ converter circuit, a wTron, and superconducting meander structures, all monolithically integrated on a single chip. (c) A zoomed-in view of the choke-channel region of a wTron; the choke width is 1.5 $\mu$m and the channel width is 3 $\mu$m. (d) The channel switching current ($I_{sw,ch}$) of the wTron was measured as the gate current ($I_{gate}$) was varied. Measurements were performed at 3.8 K. \vspace{-3mm}}
\label{fig:Fig4}
\end{figure} 

In this context, we introduce the micrometer-wide superconducting wire-based cryotron (wTron), which is a photolithography-compatible, micrometer-scale version of the nTron. The geometrical shape of the wTron is similar to that of the nTron, but the minimum feature size has been increased to 1 $\mu$m to make it compatible with standard photolithography. Although it is possible to fabricate sub-micrometer features using state-of-the-art photolithography tools, our goal in this work was to characterize wTrons fabricated using a conventional and widely accessible photolithography process. Hence, the wTron can be viewed as a photolithography-compatible counterpart to the nTron, designed to facilitate integration and scalability in practical superconductor electronics.

From the DC IV measurements, we characterized wTrons as having a similar choke-induced suppression of the channel-switching current as previously observed in nTrons \cite{R9,R31}. This suggests that the wTron can serve as a readout interface by sensing and amplifying the pulses fed to its choke. Although the wTron, with a micrometer-wide choke, will not be as sensitive to input pulses as the nTron with its sub-100 nm choke, this limitation can be addressed by adding a constant DC offset or performing pre-amplification to the input signal to match the switching current of the wTron choke. One particularly interesting application might be replacing the SQUID-based readout interfaces in SFQ and AQFP circuits with wTrons, which would reduce the footprint of the SFQ and AQFP circuits and potentially improve the reliability of their readout operations. The process node of MIT Lincoln Laboratory (MIT LL)'s SFQ7ee foundry for advanced SFQ integrated circuits is 250 nm \cite{mitll_superconducting, tolpygo2016advanced}, which can facilitate on-chip integration of wTrons with JJ-based electronics such as SFQ and AQFP circuits.

\highlightblue{Figure \ref{fig:Fig4}(a) shows an optical micrograph of a chip fabricated using MIT LL's SFQ5ee process node. The process employs photolithography to achieve feature sizes down to 350 nm and JJ diameters of 700 nm. The process includes eight niobium layers (each 200 nm thick, with $I_{sw}$ = 20 mA for 500 nm line width), one high-kinetic-inductance layer consisting of \si{MoN_x} (40 nm thick, with $I_{sw}$ = 0.5 mA for 1 $\mu$m line width and sheet inductance $L_{k}$ = 8$\pm$1 pH/sq), and two metal layers used for resistors and chip contact metallization. Interconnects between these layers are formed by etched vias filled with niobium. As shown in Figure \ref{fig:Fig4}(b), the chip contains multiple wTron devices and superconducting wire meanders in the high-kinetic-inductance layer, along with JJ-based electronic components, such as DC-to-SFQ converters and AQFP circuits, in the niobium layers.} 

\highlightblue{The wTrons and superconducting wire meanders were placed in the high-kinetic-inductance layer to take advantage of its higher kinetic inductance compared to the niobium layers, enabling a more compact footprint. Additionally, its lower switching current density helps reduce the switching energy of the wTrons. Although the SFQ5ee process allows for feature sizes as small as 350 nm, the wTrons fabricated in this chip had minimum feature sizes in the micrometer range. Figure \ref{fig:Fig4}(c) shows a fabricated wTron with 1.5 $\mu$m-wide choke and 3 $\mu$m-wide channel, while Figure \ref{fig:Fig4}(d) presents the gate-current-induced channel switching current ($I_{sw,ch}$) of the device. We observe similar channel current suppression by the choke as in Figure \ref{fig:Fig1}(c); hence, the channel switching current in Figure \ref{fig:Fig4}(d) can be empirically fitted using Equation \ref{eq:eq1}. However, the empirical values of $\alpha$ and $\beta$ for this device may differ from the previously discussed wTrons in Figure \ref{fig:Fig1}(c), as these parameters can be influenced by fabrication-specific factors, such as the superconducting material, film thickness, and lithography-induced line-edge roughness. Nonetheless, this chip tapeout represents an initial effort to demonstrate superconducting wire-based cryotron devices in the MIT LL SFQ5ee process and opens the door to the development of more complex and functional cryotron-based electronics using a superconducting foundry platform. In addition, while the JJ circuits on the chip were fabricated as part of a separate project, their presence together with wTrons on the same chip highlights promising opportunities for monolithic integration of cryotrons with JJ-based electronics using a standard superconducting foundry process.}

There has been growing interest in scaling up the detection area of SNSPDs to wafer-scale by leveraging photolithography-patterned, micrometer-wide superconducting meanders, as recent works have demonstrated single-photon sensitivity in micrometer-wide superconducting wires \cite{R18}. As nTrons have shown their merit in SNSPD readout and in scaling SNSPDs to multi-pixel arrays, wTrons can also be monolithically fabricated with superconducting micrometer-wide wire single-photon detectors (SMSPDs) using photolithography and be used for multi-pixel array readout, thereby scaling SMSPDs to a wafer-scale footprint. In parallel, there is growing interest within the superconductor electronics community in achieving integration with cryo-CMOS functionality, as it offers scalability and complexity beyond the current capabilities of nanocryotrons and JJ-based electronics. To explore this with the wTron, in this work, we demonstrated wTron driving room-temperature MOSFETs and LEDs, and discussed the circuit design parameters for wTron driving highly capacitive loads. Although, to our knowledge, there has not yet been an experimental demonstration of a superconducting wire-based cryotron interfacing with cryo-CMOS, our present work demonstrates its viability. Moreover, we presented a simulation study to optimize the capacitive load-driving performance of the wTron. We anticipate that the SPICE electrothermal model of the wTron will enable the design of highly optimized hybrid cryo-ASICs, incorporating both JJ-based electronics and cryo-CMOS, for large-scale integration in the superconducting computing ecosystem.

% Thus, the photolithography compatibility of the wTron will facilitate the on-chip integration of cryotrons with JJ-based electronics and encourage the widespread adoption of this superconducting wire-based electronics. 

\vspace{-6mm}
\section{Conclusion} 
\vspace{-3mm}
In conclusion, we demonstrated a photolithographically fabricated three-terminal superconducting switch having mA-range switching current and high output impedance. The output impedance of this device is set by the resistive hotspot in the channel, which can reach several k$\Omega$ depending on the channel width and bias current. We then discussed the optimal design considerations for driving highly capacitive CMOS loads using wTrons. We showed that a higher output impedance in the wTron is crucial for maximizing the output voltage and minimizing the turn-on delay of highly capacitive loads, such as CMOS transistors and electro-optic modulators. In addition, the micrometer-wide feature of this switch can be fabricated using less expensive photolithography tools, making it a foundry-compatible version of the nanocryotron. \highlightblue{To demonstrate this foundry compatibility, we characterized wTron devices fabricated using MIT Lincoln Laboratory’s SFQ5ee process, thereby highlighting the potential for monolithic integration with JJ-based circuits in standard superconducting foundries.} Overall, this photolithography-compatible superconducting switch paves a viable path for a superconductor–semiconductor interface, which has the potential to enable hybrid electronics with functionalities useful for various superconducting and quantum applications.

\vspace{-6mm}
\section*{Acknowledgments}
\vspace{-5mm}
This work was carried out in part through the use of MIT.nano's facilities, and the authors would like to thank M. Mondol and J. Daley of the MIT.nano laboratory for their technical support. The authors would also like to thank Owen Medeiros, Evan Golden, Alejandro Simon, and Phillip D. Keathley for their feedback during the preparation of the manuscript. We also thank Neel Parmar, David Russo, Camron Blackburn, and Andrew Wagner for their guidance and assistance in preparing the chip layout for the MIT LL SFQ5ee foundry process. The initial stage of this research was supported by the U.S. Army Research Office (ARO) and carried out under Cooperative Agreement No. W911NF-21-2-0041. The final stage of this research was supported by the National Science Foundation under Grant No. EEC-1941583, by MIT Lincoln Laboratory under the SNSPD-SFQ Line program, and by NASA under the ROSES-APRA program.

\vspace{-6mm}
\section*{Data Availability}
\vspace{-5mm}
The data that support the findings of this article are openly available in Ref. \cite{R36}. Further details are available from the authors upon reasonable request.

% \newpage

\renewcommand{\thefigure}{A\arabic{figure}}
\setcounter{figure}{0}                      
\renewcommand{\theequation}{A\arabic{equation}}
\setcounter{equation}{0}

\begin{figure}[!b]
\centering
\vspace{-3mm}
\includegraphics[width=0.7\linewidth]{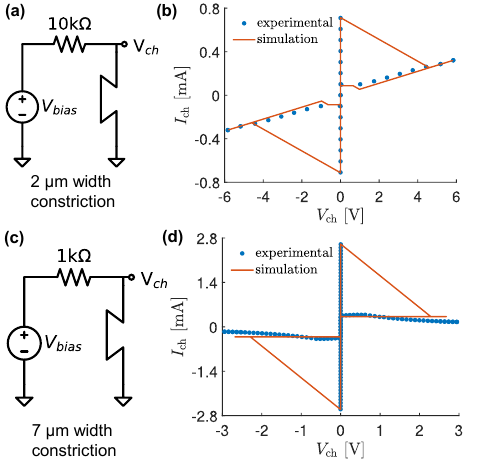}
\vspace{-5mm}
\caption{\footnotesize Benchmarking LTspice simulation results with experimental data. Circuit schematics of DC I-V measurement of (a) 2 $\mu$m-wide and (c) 7 $\mu$m-wide constrictions are shown. The corresponding I-V simulation data were compared with the experimental data in (b) and (d). \vspace{-3mm}}
\label{fig:FigA1}
\end{figure}

\begin{figure}[!b]
\centering
\vspace{-3mm}
\includegraphics[width=0.7\linewidth]{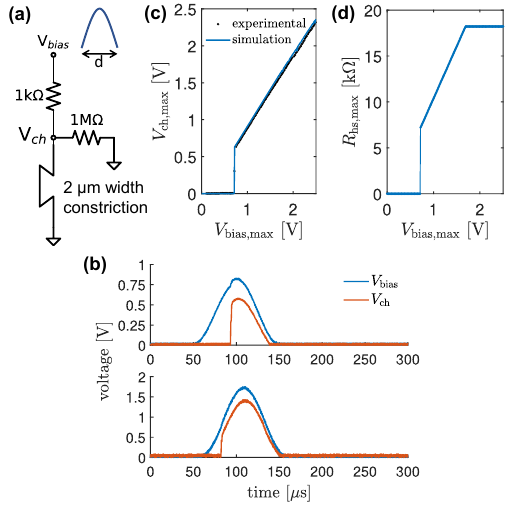}
\vspace{-5mm}
\caption{\footnotesize (a) Measurement of hot-spot-induced voltage formation in a micrometer-wide constriction. (a) A schematic illustrating biasing of a 2 $\mu$m-wide constriction with a half-cycle sinusoidal pulse. The pulse duration and amplitude were controlled with an arbitrary waveform generator (AWG). (b) Voltage across the constriction ($V_{ch}$) is plotted for a bias pulse ($V_{bias}$) having 100 $\mu$s pulse duration. Voltage appeared across $V_{ch}$ when the constriction switched, and the value of $V_{ch}$ increased with the amplitude of $V_{bias}$. (c) The amplitude of $V_{ch}$ as a function of the amplitude of $V_{bias}$ was measured and plotted in circles. Simulation data (lines) from the LTspice model are shown for comparison. (d) Simulated values of the maximum instantaneous hot-spot resistance ($R_{hs}$) formed in the constriction are shown as a function of the maximum value of the bias voltage. \vspace{-3mm}}
\label{fig:FigA2}
\end{figure}

\begin{figure}[!b]
\centering
\vspace{-3mm}
\includegraphics[width=0.7\linewidth]{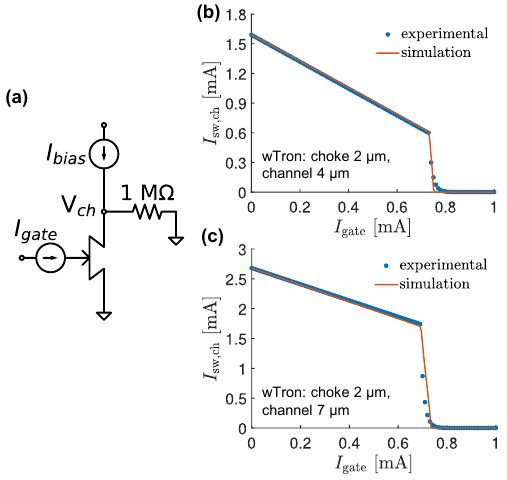}
\vspace{-5mm}
\caption{\footnotesize (a) A Schematic of DC current biasing of a wTron. (b, c) The gate-induced channel switching current ($I_{sw,ch}$) is plotted as a function of gate current and compared with LTspice simulation results for wTrons with specified choke and channel widths. \vspace{-3mm}}
\label{fig:FigA3}
\end{figure}

\vspace{-3mm}
\section*{Appendix A. Electro-thermal model of micrometer-wide superconducting wire}
\label{app:A}
\vspace{-3mm}
In this work, we used the SPICE electrothermal model of micrometer-wide superconducting cryotrons for the simulations shown in Figure \ref{fig:Fig3}. The model was originally developed for superconducting nanowire single-photon detectors (SNSPDs) \cite{R25}. To assess the validity of the model for micrometer-wide superconducting wires, we conducted benchmarking by comparing the electrothermal simulation results with experimental data. In our first experiment, we measured the DC I-V characteristics of 2 $\mu$m and 7 $\mu$m width constrictions at 3.8 K. To operate the 7 $\mu$m-wide constriction within the maximum voltage range of the voltage source (Yokogawa GS200), the bias resistance needed to be reduced to 1 k$\Omega$. In the LTspice simulation, we updated the model parameters according to the experimental device, including the material thickness, sheet resistance, critical temperature, operating temperature, constriction width, and length. The material's thermal parameters were kept unchanged, since the model was developed originally for NbN on the \si{SiO_2} substrate. A comparison of the simulation and experimental I-V data for the devices is presented in Figure \ref{fig:FigA1}. In this simulation, the switching current served as an adjustable parameter as we tuned the critical current density in the model to match the experimental results. However, the simulation model accurately predicted both the retrapping current and the load line, indicating its validity for predicting hot-spot growth in micrometer-wide wires.

After benchmarking the steady-state hot-spot resistance in micrometer-wide constrictions using DC I-V measurements, we performed transient measurements to compare the instantaneous voltage across the constriction with our LTspice simulation results. As shown in Figure \ref{fig:FigA2}(a), a half-cycle sinusoidal voltage was applied to a 2 $\mu$m-wide constriction from an arbitrary waveform generator. The 1 M$\Omega$ resistance in parallel with the constriction represents the oscilloscope in our measurement setup, as its internal impedance was configured to 1 M$\Omega$. This ensured that the resistive region in the constriction expanded to its full extent for the given bias pulse amplitude and duration. Figure \ref{fig:FigA2}(b) shows the oscilloscope traces of $V_{ch}$ for two different $V_{bias}$ amplitudes. We incrementally increased the amplitude of $V_{bias}$ and measured the corresponding amplitude of $V_{ch}$, as shown in Figure \ref{fig:FigA2}(c). Next, we performed LTspice simulations of the measured device without changing our simulation model from the previous section, and the simulated $V_{ch}$ agreed closely with the experimental data. The simulated values of the maximum instantaneous hot-spot resistance ($R_{hs,max}$) formed in the constriction are presented in Figure \ref{fig:FigA2}(d) as a function of the bias pulse amplitudes. The values of $R_{hs,max}$ reach saturation when the resistive hot-spot domain expands along the entire length of the wire. Although the voltage pulse duration in this experiment was fixed at 100 $\mu$s, the results indicate that the LTspice model can effectively capture the hot-spot expansion dynamics in micrometer-wide wires.

Next, we performed LTspice simulations of two sets of wTrons with different choke and channel widths and compared them with the DC measurement results previously presented in Figure \ref{fig:Fig1}. The cryotron LTspice model is based on the same electrothermal model used for superconducting nanowires \cite{R25}, and this wTron SPICE model is available for download from \cite{R35,R36}. The SPICE model of the nanowire-based cryotron has proven valuable in designing complex nTron-based circuits over recent years, and in this work, we aimed to extend its applicability to micrometer-wide cryotrons. Hence, we updated our device dimensions in the nTron SPICE model while keeping the thermal parameters unchanged. In addition, we introduced a parameter, $\alpha$, in the wTron SPICE model to account for the current crowding effect at the choke-channel interface and adjusted the parameter $\beta$ to match the suppression of channel switching current resulting from $I_{gate}$. The values of $\alpha$ and $\beta$ were chosen to be similar to the empirically fitted values discussed in Figure \ref{fig:Fig1}. While these values have yet to be validated across a wide range of micrometer-wide choke and channel widths for wTrons, Figure \ref{fig:FigA3} shows that it is possible to reasonably capture the I-V characteristics of micrometer-wide cryotrons in LTspice simulations by adjusting these parameters.

\renewcommand{\thefigure}{B\arabic{figure}}
\setcounter{figure}{0}                      % Reset figure counter
\renewcommand{\theequation}{B\arabic{equation}}
\setcounter{equation}{0}                    % Reset equation counter

\vspace{-3mm}
\section*{Appendix B. Analytical model of hot-spot resistance in \lowercase{w}T\lowercase{ron} with a capacitive shunt}
\label{app:B}
\begin{figure}[!b]
\centering
\vspace{-3mm}
\includegraphics[width=1\textwidth]{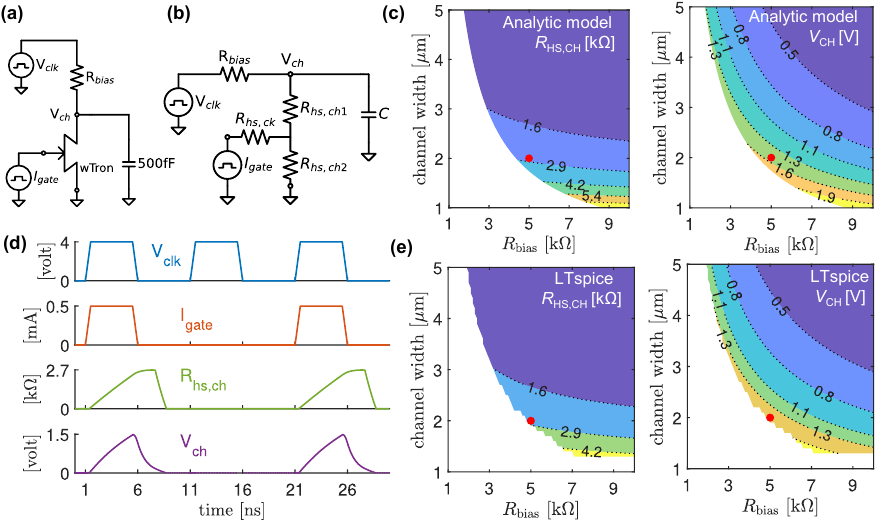}
\vspace{-12mm}
\caption{\footnotesize (a) Schematic of a wTron driving 500 fF capacitive load. (b) Equivalent circuit corresponding to (a), showing the instantaneous hot-spot resistance ($R_{hs}$) in the choke and channel regions of the wTron when it is in the resistive state. (c) The maximum value of the channel hot-spot resistance and the maximum channel voltage were calculated using the analytic model for different $R_{bias}$ and channel wire widths. In the equation, we considered $V_{clk}$ amplitude of 4 V and $I_{gate}$ amplitude of 0.5 mA, along with their pulse width of 4.5 ns, rise and fall time of 500 ps. The white region in the contour plots indicates when the current from $V_{clk}$ is sufficiently high to switch the channel wire without the presence of $I_{gate}$. This operation is not allowed, as $I_{gate}$ is intended to control the switching of the wTron. (d) LTspice simulation result of the circuit in (a), for $R_{bias}$ of 5 k$\Omega$, and wTron channel and choke width of 2 $\mu$m and 1 $\mu$m, respectively. The switching current of the choke is 0.46 mA; hence, $I_{gate}$-induced channel switching occurs in the wTron. Here, $R_{hs,ch}$ is the sum of $R_{hs,ch1}$ and $R_{hs,ch2}$. (e) The maximum value of the channel hot-spot resistance and the maximum channel voltage were calculated using the electro-thermal LTspice simulation for different channel wire widths and $R_{bias}$. The same material parameters were used in both the analytical model and the LTspice simulation. The red circles in these contour plots represent the $R_{HS,CH}$ and $V_{CH}$ value taken from (d). \vspace{-3mm}}
\label{fig:FigB1}
\end{figure}

\vspace{-3mm}
The voltage across the capacitor shunted with a wTron depends on several factors, including the channel hot-spot resistance ($R_{hs,ch}$), bias resistance ($R_{bias}$), bias voltage ($V_{clk}$), and gate current ($I_{gate}$). While it is possible to perform electro-thermal LTspice simulations and vary the circuit parameters to achieve the desired $V_{ch}$, having an analytical model to approximate the hot-spot resistance and $V_{ch}$ can facilitate this iteration process. In Figure \ref{fig:FigB1}(a), we show the schematic of a wTron shunted with a 500 fF capacitor, while Figure \ref{fig:FigB1}(b) shows its equivalent circuit when the wTron is in the resistive state. From this circuit, we can derive the maximum voltage across the capacitor using the following equation:
\begin{equation}
V_{CH} = \frac{R_{HS,CH}}{R_{HS,CH}+R_{bias}} V_{clk} + \frac{R_{HS,CH2} \times R_{bias}}{R_{HS,CH}+R_{bias}} I_{gate}
\label{eq:B1}
\end{equation}
where $R_{HS,CH}$ denotes the maximum value of the instantaneous hot-spot resistance $R_{hs,ch}$, which is the sum of $R_{hs,ch1}$ and $R_{hs,ch2}$. For a symmetric channel wTron, we can assume $R_{hs,ch1}$ to be equal to $R_{hs,ch2}$. This equation shows that having an analytical expression to estimate $R_{HS,CH}$ for a given $V_{bias}$ and $I_{gate}$ will allow us to calculate $V_{CH}$.

Following the phenomenological model for the superconducting-normal phase boundary velocity \cite{R20}, the time derivative of the hot-spot resistance can be expressed as follows:
\begin{equation}
\frac{d R_{hs}}{dt} = \frac{R_{sheet}}{w} v_{hs}
\end{equation}
where $R_{sheet}$ is the sheet resistance of the superconducting film, $w$ is the channel width of the wTron, and $v_{hs}$ is the hot-spot growth rate. The rate of hot-spot expansion depends on the current ($i$) flowing through the wire and is given by \cite{R20}
\begin{equation}
v_{hs} = 2v_{0}\frac{\psi(i/I_{sw})^2-2}{\sqrt{\psi(i/I_{sw})^2-1}} 
\end{equation}
where $v_{0} = \left(\sqrt{h_c\kappa/d}\right)/c$ is a characteristic velocity, and $\psi =  \rho I_{sw}^2/h_c w^2 d(T_C - T_S)$ is the Stekly parameter. Besides, for thin-film NbN, we can consider a linear dependence of $v_{hs}$ on the current ($i$) flowing through the wire \cite{R25}, as follows
\begin{equation}
v_{hs} = 2v_{0}(i/I_{sw}) \sqrt{\psi}
\end{equation}
If the pulse width of $V_{clk}$ and $I_{gate}$ is denoted by $T$, $R_{sheet}$ can be written as 
\begin{align}
R_{HS} &= \int_{0}^{T} \frac{d R_{hs}}{dt} dt \\
       &= \frac{R_{sheet}}{wI_{sw}} 2v_{0} \sqrt{\psi} \int_{0}^{T} i dt
\end{align}
Here, $I_{sw}$ represents the channel switching current, which can be calculated by multiplying the critical current density of the film by the cross-sectional area of the channel wire. However, the current in the channel wire varies with time as the channel hot-spot resistance ($R_{hs,ch}$) and the voltage across the channel ($V_{ch}$) change over time. Nonetheless, for a capacitively shunted wire, we can consider the capacitor to behave as an open circuit in the steady state. Hence, the current through the channel can be approximated as \(i=V_{clk}/(R_{bias}+R_{HS}) + I_{gate}\), which gives us the following equation:
\begin{equation}
R_{HS} = \frac{R_{sheet}}{wI_{sw}} 2v_{0} T (\frac{V_{clk}}{R_{bias}+R_{HS}} + I_{gate}) \sqrt{\psi}
\label{eq:B7}
\end{equation}
Rearranging this equation yields a quadratic equation for $R_{HS}$, which can be solved for given values of $V_{clk}$, $R_{bias}$, $I_{gate}$, and channel width. To compare the validity of our analytical solution with the electrothermal LTspice simulation, we considered an example of driving a 500 fF capacitor with a wTron. A periodic square wave of 100 MHz was applied to the wTron channel in series with $R_{bias}$, and the bias amplitude ($V_{clk}$) was set to 4 V.

To drive the voltage across the capacitor to a certain value, the wTron has to be in the resistive state, and we applied a gate current with 0.5 mA amplitude to control this switching. Here, the choke and channel widths of the wTron have to be selected such that the wTron becomes resistive only when $I_{gate}$ exceeds the choke switching current. Assuming a 10-nm-thick NbN film with a critical current density of 46 \si{GA/m^2}, the switching current of a 1 $\mu$m-wide wire would be 0.46 mA. Hence, we considered a choke width of 1 $\mu$m for the wTron, so that $I_{gate}$ could switch the choke and suppress the channel switching current below the bias current. This eventually initiated the expansion of a resistive hot-spot in the channel and the development of voltage across the capacitor.

For the given $V_{clk}$, $I_{gate}$, and choke width, we varied the channel width and $R_{bias}$ and calculated $R_{HS,CH}$ using Equation \ref{eq:eq2} and $V_{CH}$ using Equation \ref{eq:B1}, as plotted in Figure \ref{fig:FigB1}(c). These contour plots were useful for selecting the appropriate channel width and $R_{bias}$ to achieve the desired $V_{CH}$ across the capacitor. In addition, the plots indicate the forbidden regions of $R_{bias}$ and channel width values, which are bounded by the white area. These values of $R_{bias}$, for a 4 V amplitude $V_{bias}$, result in a bias current higher than the switching current of the corresponding channel widths in the plot, leading to loss of control over wTron switching by $I_{gate}$.

Next, we performed an LTspice simulation of the circuit with $R_{bias}$ set to 5 k$\Omega$ and the channel width set to 2 $\mu$m. The simulated values of $R_{HS,CH}$ and $V_{CH}$ were 2.69 k$\Omega$ and 1.48 V, respectively, as shown in Figure \ref{fig:FigB1}(d). For the same circuit parameters, the values of $R_{HS,CH}$ and $V_{CH}$ obtained from our analytical equations were 2.73 k$\Omega$ and 1.55 V, respectively, as indicated by the red circles in Figure \ref{fig:FigB1}(c).

Next, we computed contour plots of $R_{HS,CH}$ and $V_{CH}$ for varying $R_{bias}$ and channel width using LTspice simulations, as shown in Figure \ref{fig:FigB1}(e). The overall trend in the variation of $R_{HS,CH}$ and $V_{CH}$ as a function of $R_{bias}$ and channel width in these plots agreed reasonably well with Figure \ref{fig:FigB1}(c). This demonstrates that our analytical model provides a good approximate solution for electrothermal LTspice simulations, with significantly less computational cost. For comparison, it took nearly 3 hours to generate the data for Figure \ref{fig:FigB1}(e), whereas the data for Figure \ref{fig:FigB1}(c) were generated in a few seconds using the same computer. Additionally, the sweep interval of $R_{bias}$ and channel width was much sparser in Figure \ref{fig:FigB1}(e) than in Figure \ref{fig:FigB1}(c), as evidenced by the rough boundary edges of the white regions in Figure \ref{fig:FigB1}(e).

Although Figure \ref{fig:FigB1}(e) shows a similar white region at lower $R_{bias}$ values as in the analytical model, we observe that it extends across all values of $R_{bias}$ for channel widths below 1.3 $\mu$m. As we know, the RC time constant of the circuit is given by $(R_{HS,CH} || R_{bias})C$, which should be small enough to allow the restoration of superconductivity in the wTron channel before the subsequent $V_{clk}$ pulses. Otherwise, the wTron will latch and remain resistive for the entire periodic $V_{clk}$ input. In Figure \ref{fig:FigB1}(c), we can see that the values of $R_{HS,CH}$ are higher than 5.4 k$\Omega$ for channel widths below 1.3 $\mu$m, corresponding to a larger RC time constant for the wTron to reset for a 100 MHz $V_{clk}$. Hence, those values of $R_{bias}$ for the corresponding channel width were not allowed in our design. Although this effect could be incorporated into our analytical model, we chose to ignore it for simplicity. Nonetheless, our objective to use the analytical model for rapid parameter sweeps remains promising in facilitating design iteration and improving our understanding of the tradeoffs in wTron circuits driving capacitive loads.

% \section*{References}
\vspace{-3mm}
% \bibliography{wTron_Ref}{}

\bibliographystyle{ieeetr}

\end{document}